# Investigating End-user Acceptance of Last-mile Delivery by Autonomous Vehicles in the United States

Antonios Saravanos[1]([✉]) [0000-0002-6745-810X], Olivia Verni[1], Ian Moore[1],
Sall Aboubacar[1], Jen Arriaza[1], Sabrina Jivani[1], Audrey Bennett[1],
Siqi Li[1], Dongnanzi Zheng[1], Stavros Zervoudakis[1]

[1] New York University, New York NY, USA
```
{saravanos, orv209, inm220, as15872, jen.arriaza, sj2903,
     aab883, sl7137, dz40, zervoudakis}@nyu.edu
```

**Abstract.** This paper investigates the end-user acceptance of last-mile delivery carried out by autonomous vehicles within the United States. A total of 296 participants were presented with information on this technology and then asked to complete a questionnaire on their perceptions to gauge their behavioral intention concerning acceptance. Structural equation modeling of the partial least squares flavor (PLS-SEM) was employed to analyze the collected data. The results indicated that the perceived usefulness of the technology played the greatest role in end-user acceptance decisions, followed by the influence of others, and then the enjoyment received by interacting with the technology. Furthermore, the perception of risk associated with using autonomous delivery vehicles for last-mile delivery led to a decrease in acceptance. However, most participants did not perceive the use of this technology to be risky. The paper concludes by summarizing the implications our findings have on the respective stakeholders, and proposing the next steps in this area of research.

**Keywords:** technology adoption, end-user acceptance, last-mile delivery, autonomous delivery vehicles, autonomous delivery robots.

## 1      Introduction

In this work, we investigate the end-user acceptance of last-mile delivery carried out by autonomous delivery vehicles (ADVs), also known as autonomous delivery robots (ADRs), within the United States. It has been noted that "the rapid growth of e-commerce and package deliveries across the globe is demanding new solutions to meet customers' desire for more and faster deliveries" [8]. Accordingly, "with the significant rise in demand for same-day instant deliveries, several courier services are exploring alternatives to transport packages in a cost- and time-effective, as well as, sustainable manner" [23]. Thus, an understanding of the determinants leading to the acceptance of this technology by end-users is of value to stakeholders. The first study to examine such acceptance was carried out by Kapser and Abdelrahman [15], who focused exclusively on the German market. However, considerable evidence demonstrates that consumer behavior concerning the adoption of technology varies by culture [2, 3, 7, 13, 25, 27].

Consequently, there is value in expanding that work to investigate the United States consumer context.

## 2      Materials and Method

For our work, we rely on Kapser and Abdelrahman's [15] technology adoption model – an extension of the Unified Theory of Acceptance and Use of Technology (hereafter UTATU2) model, which has incorporated consumer risk perception. The model contains a total of eight constructs with respect to the 'Behavioral Intention' (BI) factor: 'Effort Expectancy' (EE), 'Facilitating Conditions' (FC), 'Hedonic Motivation' (HM), 'Performance Expectancy' (PE), 'Perceived Risk' (PR), 'Price Sensitivity' (PS), 'Social Influence' (SI), and 'Trust in Technology' (TT). It also contains six control variables: 'Age' (AGE), 'Education' (EDU), 'Employment' (EMP), 'Gender' (GEN), 'Heard Before' (HEB), this factor reflecting whether an individual has previously heard of the autonomous delivery vehicle technology, and 'Income' (INC). For conciseness, we refer the reader to Kapser and Abdelrahman's [15] paper for a more complete definition of each of these constructs.

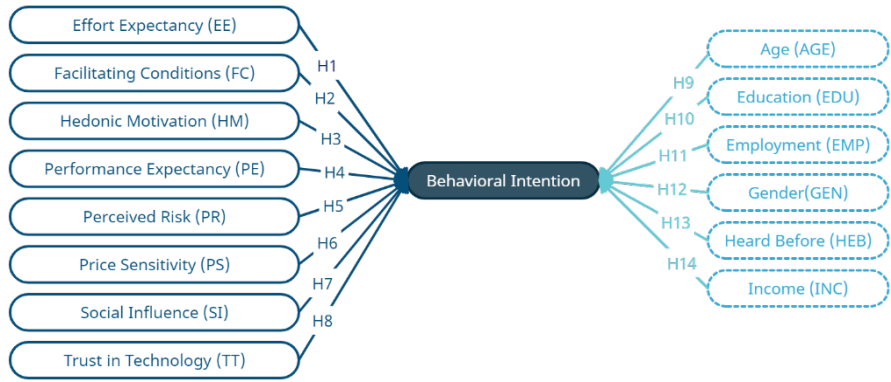

**Fig. 1.** The proposed model adapted from Kapser and Abdelrahman [15], who based their work on Venkatesh et al.'s [13] UTAUT2 framework.

We supplement this model by incorporating a factor to reflect consumer trust in the technology, appropriately titled 'Trust in Technology' (TT), adapting the construct from the research of Kim et al. [16]. This inclusion is supported by the work of Panagiotopoulos and Dimitrakopoulos [20], who write [9] that "few adaptations of the technology acceptance model have considered trust as a determinant of acceptance; however, those who have done so have found trust to be a determinant of intention to use, i.e. in the context of e-services and e-government applications", pointing to the publications of Mou et al. [19] and Gupta et al. [9]. Furthermore, in a later study by Kapser et al. [14], also investigating autonomous delivery vehicles, the authors incorporate such a construct. Lastly, it should be noted that the practice of adapting the UTAUT2 model for a specific technology is quite common, and Hino [11] describes the (UTAUT) model as "flexible enough to be modified by incorporating additional variables into the original model". The author goes on to point to the work of Yu [26], who justifies this practice by writing that the model is "thus better explaining the acceptance of innovative technology". The final model can be seen illustrated in Fig. 1, and the corresponding hypotheses are outlined in Table 1.

**Table 1.** List of Hypotheses with Respective Relationships.

| Hypothesis | Relationship |
| --- | --- |
| H1 | Effort Expectancy (EE) → Behavioral Intention (BI) |
| H2 | Facilitating Conditions (FC) → Behavioral Intention (BI) |
| H3 | Hedonic Motivation (HM) → Behavioral Intention (BI) |
| H4 | Performance Expectancy (PE) → Behavioral Intention (BI) |
| H5 | Perceived Risk (PR) → Behavioral Intention (BI) |
| H6 | Price Sensitivity (PS) → Behavioral Intention (BI) |
| H7 | Social Influence (SI) → Behavioral Intention (BI) |
| H8 | Trust in Technology (TT) → Behavioral Intention (BI) |
| H9 | Age (AGE) → Behavioral Intention (BI) |
| H10 | Education (EDU) → Behavioral Intention (BI) |
| H11 | Employment (EMP) → Behavioral Intention (BI) |
| H12 | Gender (GEN) → Behavioral Intention (BI) |
| H13 | Heard Before (HEB) → Behavioral Intention (BI) |
| H14 | Income (INC) → Behavioral Intention (BI) |

An experimental approach was taken to evaluate the aforementioned model (see Fig. 1), following the specifications also set by Kasper and Abdelrahman [15] for their study, namely: an information sheet to expose participants to the autonomous delivery vehicle technology; subsequently an instrument (i.e., questionnaire) designed to capture participant demographics and participants' perception of the technology. The questionnaire was slightly adapted for the United States context with regard to language and units of measurement. We would note that the questions for the construct of 'Trust in Technology' were derived from the work of Kim et al. [16]. All questions for the constructs were measured using a 7-point Likert scale (strongly disagree; moderately disagree; somewhat disagree; neutral, neither agree nor disagree; somewhat agree; moderately agree; and strongly agree).

### 2.1 Sample and Data Collection

A series of web-based experiments were performed where participants were solicited using the Amazon Mechanical Turk crowdsourcing platform, which has been demonstrated to be an effective tool [22] to solicit participants but requires care [24] with respect to participant inattention. Following the obtainment of informed consent, participants were asked to read through the aforementioned 'informational sheet' on autonomous delivery vehicles, and then complete the questionnaire. A total of 296 participations were collected. Specifically, participants were asked to 'take part in a research study soliciting perceptions on autonomous delivery vehicles' and were also presented with two keywords: 'experiment' and 'user perceptions'. The selected qualification requirements for all requesters on the portal were: (1) a HIT Approval Rate greater than 98; (2) that they are located in the United States; (3) a Number of HITs Approved greater than 5000. The survey included two attention check questions adopted from Abbey and Meloy [1]. The first was 'I would rather eat a piece of paper than a piece of fruit', and the second 'At some point in my life, I have had to consume water in some form'. Participants were asked to answer these questions through a 7-point Likert scale (again ranging from 'strongly disagree' to 'strongly agree'). Accordingly, participants who did not answer the first question with 'strongly disagree' and the second with 'strongly agree' were removed from the sample as they were undoubtedly not paying attention. This resulted in a total of 34 participants who failed these attention checks being removed, leaving 262 participants. The sample characteristics can be seen in more detail in Table 2.

Table 2. Participant Demographics.

| Characteristic | Category | N | Percentage |
|---|---|---|---|
| Age | 18-25 | 1 | 0.38% |
| | 26-30 | 1 | 0.38% |
| | 31-35 | 24 | 9.16% |
| | 36-45 | 52 | 19.85% |
| | 46-55 | 95 | 36.26% |
| | 56 or older | 43 | 16.41% |
| | Prefer not to answer | 46 | 17.55% |
| Gender | Female | 105 | 40.08% |
| | Male | 157 | 59.92% |
| Income | Less than $10,000 | 4 | 1.53% |
| | $10,000 - $19,999 | 8 | 3.05% |
| | $20,000 - $29,999 | 12 | 4.58% |
| | $30,000 - $39,999 | 24 | 9.16% |
| | $40,000 - $49,999 | 30 | 11.45% |
| | $50,000 - $59,999 | 36 | 13.74% |
| | $60,000 - $69,999 | 22 | 8.40% |
| | $70,000 - $79,999 | 22 | 8.40% |
| | $80,000 - $89,999 | 21 | 8.02% |
| | $90,000 - $99,999 | 18 | 6.87% |
| | $100,000 - $149,999 | 15 | 5.73% |
| | $150,000 or more | 34 | 12.98% |
| | Prefer not to answer | 16 | 6.11% |
| Schooling | High school graduate (high school diploma or equivalent including GED) | 1 | 0.38% |
| | Some college but no degree | 32 | 12.21% |
| | Associate degree in college (2-year) | 42 | 16.03% |
| | Bachelor's degree in college (4-year) | 37 | 14.12% |
| | Master's degree (e.g., MA, MS) | 115 | 43.89% |
| | Professional degree (e.g., MBA, MFA, JD, MD) | 28 | 10.69% |
| | Doctoral degree (e.g., PhD, EdD, DBA) | 4 | 1.53% |
| | Prefer not to answer | 3 | 1.15% |

## 3    Analysis and Results

The proposed model is validated using the partial least squares-structural equation modeling (PLS-SEM) approach. PLS-SEM can be described as "a causal modeling approach aimed at maximizing the explained variance of the dependent latent constructs" [10]. Hair et al. [10] explain that "a structural equation model with latent constructs has two components" and further elaborate, "the first component is the structural model—typically referred to as the inner model in the PLS-SEM context—which shows the relationships (paths) between the latent constructs". They also state that "the second component of the structural equation model comprises the measurement models, also referred to as outer models in the PLS-SEM context" [10].

With respect to the measurement model, we assessed convergent validity, construct reliability, and discriminant validity. For the first, convergent validity, we looked at the factor loadings, removing those manifest variables that had values that were lower than 0.7 (see Chin [5]). This saw us remove PS3 (0.552) and TT4 (0.698), with the remaining items being statistically significant (see Table 3). Secondly, with regard to construct reliability, we relied on composite reliability (CR) and Cronbach's Alpha, where, given that all statistics were above the 0.7 recommended threshold, we concluded that construct reliability was satisfactory (see Table 4). Next, discriminant validity was tested using the Fornell-Larcker criterion (see Table 5) and cross-loadings. All values were within the recommended guidelines; therefore, we assumed that the discriminant validity of our measurement model was satisfactory. Thus, in short, our model was found to have suitable convergent validity, construct reliability, and discriminant validity.

**Table 3.** Summary of Convergent Validity Testing.

| Factor | Item | Loading | t-Statistic | AVE |
|---|---|---|---|---|
| BI | BI1 | 0.955 | 141.305** | 0.912 |
|  | BI2 | 0.955 | 132.313** |  |
| EE | EE2 | 0.945 | 99.233** | 0.894 |
|  | EE4 | 0.947 | 106.845** |  |
| FC | FC1 | 0.813 | 21.599** | 0.645 |
|  | FC2 | 0.816 | 25.245** |  |
|  | FC3 | 0.821 | 23.734** |  |
|  | FC4 | 0.760 | 16.380** |  |
| HM | HM2 | 0.965 | 231.385** | 0.914 |
|  | HM3 | 0.947 | 90.150** |  |
| PE | PE1 | 0.929 | 74.056** | 0.886 |
|  | PE3 | 0.947 | 127.476** |  |
|  | PE4 | 0.947 | 105.055** |  |
| PR | PR2 | 0.950 | 101.295** | 0.901 |
|  | PR3 | 0.948 | 95.600** |  |
| PS | PS1 | 0.903 | 54.842** | 0.796 |
|  | PS2 | 0.919 | 62.701** |  |
|  | PS4 | 0.905 | 51.163** |  |
|  | PS5 | 0.841 | 27.826** |  |
| TT | TT1 | 0.958 | 146.783** | 0.911 |
|  | TT2 | 0.951 | 120.325** |  |

∗ $p<0.05$; ** $p < 0.01$.

**Table 4.** Summary of Reliability Testing.

| Construct | Number of Items | Cronbach's Alpha | CR |
|---|---|---|---|
| BI | 2 | 0.904 | 0.954 |
| EE | 2 | 0.882 | 0.944 |
| FC | 4 | 0.816 | 0.879 |
| HM | 2 | 0.907 | 0.955 |
| PE | 3 | 0.935 | 0.959 |
| PR | 2 | 0.890 | 0.948 |
| PS | 4 | 0.915 | 0.940 |
| TT | 2 | 0.902 | 0.953 |

**Table 5.** Fornell-Larcker Criterion

|     | BI     | EE     | FC     | HM     | PE     | PR     | PS    | SI    | TT    |
|-----|--------|--------|--------|--------|--------|--------|-------|-------|-------|
| BI  | **0.955** |        |        |        |        |        |       |       |       |
| EE  | 0.442  | **0.946** |        |        |        |        |       |       |       |
| FC  | 0.433  | 0.758  | **0.803** |        |        |        |       |       |       |
| HM  | 0.644  | 0.373  | 0.363  | **0.956** |        |        |       |       |       |
| PE  | 0.736  | 0.365  | 0.326  | 0.632  | **0.941** |        |       |       |       |
| PR  | -0.476 | -0.375 | -0.414 | -0.367 | -0.346 | **0.949** |       |       |       |
| PS  | 0.482  | 0.191  | 0.169  | 0.427  | 0.485  | -0.188 | **0.892** |       |       |
| SI  | 0.599  | 0.290  | 0.268  | 0.520  | 0.528  | -0.238 | 0.496 | **1.000** |       |
| TT  | 0.644  | 0.454  | 0.446  | 0.572  | 0.610  | -0.628 | 0.413 | 0.480 | **0.954** |

Note: The square root of AVE appears in bold type.

Regarding the structural model, we first appraised the level of collinearity of our latent variables by looking at the variance inflation factor, also known as the VIF (see Table 6). We removed all values that were above the recommended threshold of 5 (see Hair et al. [10]) to resolve any issues regarding collinearity. These were: SI1 (12.850), TT3 (11.117), PE2 (7.608), SI2 (7.233), BI3 (7.164), EE1 (6.208), HM1 (6.105), PR1 (6.074), and EE3 (5.802). With respect to the variance explained by our model, BI had an $R^2$ of 0.696 which, according to Chin [5], can be described as substantial, given that it is above 0.67. With respect to the individual factors (see Table 7), PE played the greatest role, where an increase of 1 unit in PE ($\beta = 0.388$; $p < 0.01$) resulted in an increase of 0.388 units in BI; thus, hypothesis 4 was supported. Following PE, we observe SI ($\beta = 0.179$; $p < 0.01$) playing the second most prominent role, where an increase in 1 unit in SI leads to an increase of 0.179 units in BI; thereby supporting hypothesis 7. Next was HM ($\beta = 0.162$; $p < 0.05$), where an increase of 1 unit in HM induces an increase of 0.162 units in BI, allowing us to accept hypothesis 3. Subsequently, the fourth greatest role was played by GEN ($\beta = 0.099$; $p < 0.01$), where identifying as male led to a 0.099 increase in BI, consistent with hypothesis 12. Lastly, PR ($\beta = -0.121$; $p < 0.05$) exhibited a negative effect. Accordingly, an increase of 1 unit in PR leads to a decrease of 0.121 in BI; therefore, the result was consistent with hypothesis 5. The results from the hypotheses testing are summarized in Table 8.

**Table 6.** Collinearity Statistics (VIF).

| Factor | Item | VIF |
|---|---|---|
| BI | BI1 | 3.120 |
|    | BI2 | 3.120 |
| EE | EE2 | 2.647 |
|    | EE4 | 2.647 |
| FC | FC1 | 2.078 |
|    | FC2 | 1.711 |
|    | FC3 | 1.959 |
|    | FC4 | 1.486 |
| HM | HM2 | 3.206 |
|    | HM3 | 3.206 |
| PE | PE1 | 3.223 |
|    | PE3 | 4.874 |
|    | PE4 | 4.947 |
| PR | PR2 | 2.809 |
|    | PR3 | 2.809 |
| PS | PS1 | 2.866 |
|    | PS2 | 3.718 |
|    | PS4 | 3.258 |
|    | PS5 | 2.286 |
| TT | TT1 | 3.072 |
|    | TT2 | 3.072 |

**Table 7.** Structural Model Results.

| Path | B | t-statistic |
|---|---|---|
| EE → BI | 0.034 | 0.596 |
| FC → BI | 0.079 | 1.519 |
| HM → BI | 0.162 | 2.492* |
| PE → BI | 0.388 | 5.993** |
| PR → BI | -0.121 | 2.486* |
| PS → BI | 0.051 | 1.238 |
| SI → BI | 0.179 | 3.118** |
| TT → BI | 0.080 | 1.265 |
| AGE → BI | 0.020 | 0.539 |
| EDU → BI | 0.004 | 0.112 |
| EMP → BI | 0.052 | 1.527 |
| GEN → BI | 0.099 | 2.854** |
| HEB → BI | -0.045 | 1.111 |
| INC → BI | -0.052 | 1.248 |

\* p<0.05; \*\* p < 0.01.

**Table 8.** Hypotheses Testing Results.

| Hypothesis | Relationship | Decision |
|---|---|---|
| H1 | EE → BI | Not Supported |
| H2 | FC → BI | Not Supported |
| H3 | HM → BI | Supported |
| H4 | PE → BI | Supported |
| H5 | PR → BI | Supported |
| H6 | PS → BI | Not Supported |
| H7 | SI → BI | Supported |
| H8 | TT → BI | Not Supported |
| H9 | AGE → BI | Not Supported |
| H10 | EDU → BI | Not Supported |
| H11 | EMP → BI | Not Supported |
| H12 | GEN → BI | Supported |
| H13 | HEB → BI | Not Supported |
| H14 | INC → BI | Not Supported |

## 4   Discussion and Conclusions

In this paper, we explored the end-user acceptance of last-mile delivery carried out by autonomous delivery vehicles, complementing the existing literature, where "an insufficient number of studies exist that focus explicitly on the acceptance of ADVs in the context of last-mile delivery" [15]. Moreover, Kapser and Abdelrahman [15] bring attention to the work of Hulse et al. [12], who write that "to date, there is limited research on the psychological factors that determine public acceptance of AVs from an outside vehicle perspective". While this technology holds great potential, "societal benefits will not be achieved unless these vehicles are accepted and used by a critical mass of people; thus, it will be important to understand consumers' acceptance" [20]. In particular, we built on the original work of Kapser and Abdelrahman [15] by exploring the topic from the cultural context of the United States.

Overall, consumers in the United States appear to hold a slightly favorable view of this technology, with a mean user acceptance (i.e., 'Behavioral Intention') score of 4.545 out of 7. This is in contrast to the findings of Kapser and Abdelrahman [15] who, for the German market, report that "respondents seem to hold neutral acceptance of towards the use of ADVs". Furthermore, we found that consumer perception of the perceived usefulness of this technology (i.e., 'Performance Expectancy') was the greatest determinant in consumer acceptance decisions. This finding was not surprising as it mirrors the findings of Venkatesh et al. [13], who note that "the performance expectancy construct within each individual model". Indeed, they state that it "is the strongest prediction of intention and remains significant at all points of measurement in both voluntary and mandatory settings". This finding also reaffirms the observations within the literature by Kapser and Abdelrahman [15], who write that this is "concurrent with previous AVs acceptance studies". However, one difference between our results and Kapser and Abdelrahman's [15] is with respect to the magnitude of the factor. On the contrary, they find that, for a German audience, price sensitivity is the greatest predictor of consumer acceptance of the technology, "indicating that the price for the delivery is more important to potential users than the usefulness of the technology itself". In contrast, in a United States context, price plays no statistically significant role in consumer acceptance decisions. This finding also provides insight to those offering such a service, suggesting that in the United States market a higher price could be charged, assuming that the price is within reasonable bounds, while there is an expectation that the product would give them value (i.e., be useful to them).

The second greatest predictor of end-user acceptance of last-mile delivery carried out by autonomous delivery vehicles was the opinion held by others (i.e., 'Social Influence'). Indeed, we find that "our respondents are likely to depend on their peers' opinion in regard to ADVs" [15]. The third greatest factor was the enjoyment experienced by interacting with the technology (i.e., Hedonic Motivation). This finding is again similar to what is reported by Kapser and Abdelrahman [15], who comment that "fun and entertainment derived from using ADVs seems important to determine user acceptance". Furthermore, it reflects what appears in the mainstream literature (see

Madigan et al. [17] and Moták et al. [18]). However, there is one important difference that should be mentioned. In the German context (per Kapser and Abdelrahman's [15]), enjoyment plays the second greatest effect on consumer decisions and the opinion of others the third, whereas in the United States context, the opinion of others is second and enjoyment is third. In other words, what others think is more important in consumer acceptance decisions in the United States, compared to Germany, than how much fun they have using it.

Gender was also found to influence acceptance, with males being more likely to accept the technology over females. This finding is contrary to two studies examining end-user acceptance of last-mile delivery carried out by autonomous delivery vehicles in Germany: the first by Kapser and Abdelrahman [15] and the second by Kapser et al. [14], where both studies find gender to play no statistically significant role. Lastly, we find that the perception of risk associated with the use of the technology (i.e., 'Perceived Risk') influences acceptance negatively. So, if a consumer perceives the technology to be risky, they are less likely to accept it. However, the average value for the construct was 2.675 out of 7, which indicates that, for the most part, the technology is not perceived as risky. Unexpectedly, support for using the technology did not have a statistically significant effect on consumers' behavioral intention to accept. This is in contrast to what was reported by Kapser and Abdelrahman [15], who conclude that "external resources like peer support plays an important aspect in user acceptance of ADVs". As well as what they note appears in the literature (e.g., Madigan et al. [17], Choi and Ji [6], and Buckley et al. [4]). Thus, in contrast to the German market, where consumers expect support if they are to accept this technology, in the United States market such support is not needed. Rather individuals have the expectation that they will figure it out – perhaps that is part of the enjoyment they expect to experience, or perhaps it adds a social element of working with peers to resolve the issue, giving a greater desire to figure things out on one's own.

### 4.1 Implications

Our research has both theoretical and practical implications. From the theoretical dimension, the work validates the model initially proposed by Kapser and Abdelrahman [15] for a United States cultural context. From the practical dimension, this study offers insight into the minds of the United States consumer with respect to end-user acceptance of last-mile delivery carried out by autonomous delivery vehicles. The findings can therefore inform both industry and governments as they move to promote the use of such innovative technology, affording society with a cleaner, more effective, and efficient solution to delivery. First, organizations know that the United States consumer sees the technology favorably and is concerned about quality rather than price when it comes to having access to this technology. This means that organizations should make sure that the technology is one that consumers find useful at the same time, knowing that they can charge for this service without worrying that it may lead to disenfranchisement on the part of the consumer. However, this is assuming prices are within reasonable bounds, and further research is warranted to truly understand how consumers will react to prices that go beyond such bounds. Second, consumers are influenced by the opinion of their peers when it comes to accepting this technology; consequently, "peer pressure can be taken into consideration for marketing purposes" [15]. Third, as the perception of the technology being fun to use leads to increased acceptance, efforts to gamify the experience are warranted. Fourth, those who identify as male have a greater affinity to accept the technology; this also informs marketers as they prepare campaigns. Fifth, while, for the most part, consumers do not perceive this technology as being risky, in those cases where it is perceived in that way (i.e., risky), interventions are necessary to inform. Lastly, consumers are not going to be seeking formalized support for using the technology. On the one hand, this may lead to fewer support staff; however, it does introduce other issues that need to be addressed. Specifically, efforts must be taken to help consumers that will resist contacting or interacting with technical support. It is possible that consumers appreciate the challenge of figuring out how to use such a novel technology on their own as fun. Consequently, there is a need to ensure the user interface is easy to understand and that there are alternative informal resources that can be used in lieu.

### 4.2 Limitations and Future Research

There are four limitations that should be mentioned that concurrently allow us to share ways through which this research can be further expanded in the future. The first has to do with the data for this study being collected during the Covid pandemic, which may have influenced consumer behavior during this period. This understandable change in consumer attitude is presented within the context of autonomous delivery vehicles in the work of Pani et al. [21], who write that "the ongoing COVID-19 pandemic has put a global spotlight on ADRs for contactless package deliveries". They explain that there has been "a surge in the public interest and demand for ADRs since it can provide contactless delivery, a highly sought-after service under the directives of social distancing". Pani et al. [21] conclude that "consumers, businesses, and governments have switched from being cautious beta testers into eager early adopters" [21]. Following the pandemic, it might be of value to investigate how, if at all, consumer behavior with respect to this technology changes.

The second limitation has to do with current consumer exposure to the technology being limited. Very few people have had actual access and thus experienced this technology. Hence the findings of this work are based on participants' imagination with respect to the operation of autonomous delivery vehicles and how they would carry out last-mile delivery to end-users. Subsequently, a study that exposed participants to the technology may yield different findings and would be of value to our community. The third is with respect to our study employing a generic perspective of autonomous delivery vehicles for last-mile delivery. Accordingly, brand may influence user acceptance and should be investigated in greater detail.

Lastly, we would point out that our study has presented participants only with the positive aspects of this technology. For example, with any introduction to technology that offers automation, there is the strong possibility of job loss. It is unclear how consumers would behave when knowing such tradeoffs.

**Acknowledgments.** This research was funded in part through the New York University School of Professional Studies Full-Time Faculty Professional Development Fund.